# Berreman Embedded Eigenstates for Narrowband Absorption and Thermal Emission

Zarko Sakotic[1*], Alex Krasnok[2], Norbert Cselyuszka[1,3], Nikolina Jankovic[1], and Andrea Alú[2*]

[1]*BioSense Institute-Research Institute for Information Technologies in Biosystems, University of Novi Sad, Dr Zorana Djindjica 1a, 21101, Novi Sad, Serbia*

[2]*Advanced Science Research Center, City University of New York, New York, NY 10031, USA*

[3]*Silicon Austria Labs, Sensor Systems, Microsystem Technologies, Villach, Austria*

*To whom correspondence should be addressed: zarko.sakotic@biosense.rs, aalu@gc.cuny.edu

## Abstract

Embedded eigenstates are nonradiative modes of an open structure with momentum compatible with radiation, yet characterized by unboundedly large Q-factors. Traditionally, these states originate from total destructive interference of radiation from two or more non-orthogonal modes in periodic structures. In this work, we demonstrate a novel class of embedded eigenstates based on Berreman modes in epsilon-near-zero (ENZ) layered materials and propose realistic silicon carbide (SiC) structures supporting high-Q ($\sim 10^3$) resonances based on these principles. The proposed structures demonstrate strong absorption in a narrow spectral and angular range, giving rise to quasi-coherent and highly directive thermal emission.

## Introduction

Scattering of light is a ubiquitous process, which has driven human curiosity for thousands of years, from ancient Greek philosophers to modern physicists. For a scattering process to occur, electromagnetic waves need to interact with matter. This interaction lies at the center of today's experimental physics and technology, both in classical and quantum regimes. A property intimately tied to light-matter interaction is wave confinement in the form of the system eigenmodes. The quest for confining large amounts of electromagnetic energy into small volumes has been at the forefront of technological advances in recent decades, where high Q-factors and low mode volumes have been sought-after with various approaches [1]–[3].

A recent approach to this problem holds the promise for extreme light confinement in the form of embedded eigenstates (EE) or bound states in the continuum (BIC) [3],[4],[13]-[22],[5],[23],[6]-[12]. This phenomenon has been initially introduced in quantum mechanics by Wigner and Von Neumann in 1939, who showed that a tailored potential distribution can bound an electron residing above a potential well, without decaying to the continuum [7]. Recently, the ubiquitous wave nature of this phenomenon was realized and explored in different wave domains, including mechanics [24], acoustics [25] and electromagnetics [9]. The presence of resonances in a system usually results in local-field amplification and energy decay due to radiative loss. Contrary to intuition, EEs represent eigenmode solutions of open structures that do not radiate, despite being compatible with radiation in terms of momentum. As a result, these resonant states are characterized by unbounded Q-factors, making this phenomenon of great importance for light-matter interactions [3]. Photonic EEs have become a particularly intense research direction in the last few years since the maturity of nanofabrication technologies has allowed experimental realizations of such structures in the microwave, THz, visible and infrared ranges [13]-[15],[17],[18],[26]. Trivial EEs can arise due to symmetry-forbidden radiative decay [8], usually appearing at the band edge of a periodic open structure. More recently, non-symmetry protected EEs have been engineered based on the Friedrich and Wintgen model, based on which two or more nonorthogonal modes are strongly coupled and both radiate towards the excitation channel. An EE state arises when the interference of partial waves of the modes becomes purely destructive [8].

A conceptually different approach to realize light-trapping structures that support EEs is based on using epsilon-near-zero (ENZ) materials [16],[27]-[29], which represent a versatile platform to boost light-matter interactions [30]-[33]. ENZ materials have been exploited for Purcell-effect enhancement [34]-[36], perfect absorbers [34],[37],[38], active optoelectronic devices [39], enhanced nonlinear effects [40]-[44] and to control the radiation of emitters [45],[46]. Apart from field enhancement, ENZ regimes were shown to enable peculiar optical phenomena due to inherent "wave stretching" – tailoring the radiation pattern and directivity [47], as well as supercoupling [48] and other promising applications at microwaves and optical frequencies [49]. At the bulk plasma resonance, the wave impedance in an ENZ becomes infinite [49],[50], which provides unique opportunities to engineer EEs. When the plasma resonance is combined with a geometric resonance, they can give rise to divergent Q-factors and converging scattering lines [16],[27]-[29]. ENZ responses can be found in various natural materials, such as metals (Au, Ag), transparent conductive oxides (ITO, AZO), polar dielectrics (SiC, AlN), and artificial materials [49].

In this paper, we discuss a class of EEs stemming from Berreman modes in epsilon-near-zero materials comprising dielectric layers. We show that these modes manifest themselves as zero reflection contours in the reflection spectra, associated with the complete transmission of energy for plane-wave excitation. Using the transverse resonance technique we analyze and discuss the nature of the supported modes in the considered geometries. In particular, we show that a uniform ENZ slab supports a trivial embedded eigenstate at normal incidence, in analogy to symmetry-protected EEs found in periodic structures. We then show that a three-layer system consisting of ENZ-dielectric-ENZ layers supports different orders of Berreman modes, which enable both trivial and *accidental* EEs. Next, by using the proposed concepts, we demonstrate that we can maximize the Q-factors and field enhancements of the supported modes in ENZ materials with realistic loss. We use silicon carbide (SiC) as a long wave-IR ENZ material and demonstrate quasi-EEs with high Q-factors (~$10^3$). Finally, we apply these ideas to design an extremely narrowband perfect absorber/thermal emitter based on the discussed concepts, which may provide temporally and spatially narrowband thermal emission based on embedded eigenstates [51].

## Results and Discussion

**Embedded eigenstates and Berreman modes**. To analyze the continuum of modes supported by a multilayer planar structure with ENZ properties, we employ the transverse resonance technique [52]. In this approach, a 1D waveguiding structure is represented by an equivalent transverse transmission line network, and the supported modes can be found through the resonance condition

$$Z_{\text{up}} + Z_{\text{down}} = 0, \tag{1}$$

where $Z_{\text{up}}$ and $Z_{\text{down}}$ represent impedances looking up and down towards the boundaries of the waveguiding structure from an arbitrary point in the transverse network. For closed lossless structures, these solutions can be found for real $\omega$ and real $k$. However, in addition to guided modes below the light line, open structures also support *leaky* modes with complex $\omega$ for real $k$, or complex $k$ for real $\omega$, where the imaginary parts describe the modal decay. These solutions are associated with poles of the Green's function [52],[53], and in the complex frequency representation (for fixed real $k$) they appear as poles of the S-matrix eigenvalues. Indeed, another essential theoretical tool that we use in this work is complex frequency analysis, describing the underlying phenomena through the study of poles and zeros of the scattering matrix eigenvalues/reflection, which can capture many elastic scattering phenomena relying on the analytical continuity of the *S*-matrix in the complex plane and Weierstrass factorization theorem [3],[54]-[56]. The poles of the *S*-matrix represent self-sustained solutions or *eigenmodes*, corresponding to purely outgoing waves, while the zeros represent *absorbing modes* or solutions representing solely incoming waves [55]. The distance of a pole from the real frequency axis reveals the decay rate of the corresponding mode for given real *k*, which means that complex poles of higher Q-factor modes are found closer to the real frequency axis. If a lossless, passive system supports an embedded eigenstate, i.e., a mode with zero decay rate, the *S*-matrix pole corresponding to that mode will be lying on the real frequency axis. In a passive system, this is only possible if there is also a degenerate *S*-matrix *zero* occurring at the same real frequency, where they merge and cancel out [3],[29].

We start our mode analysis from a simple planar slab of ideal ENZ material of thickness *t* = 800 nm, infinitely extended in the *x* and *y* directions, shown in the inset of Figure 1(a), consistent with the analysis in [47]. The permittivity of the slab $\varepsilon_{enz}$ is assumed to follow a Drude dispersion with vanishing loss and plasma frequency $\omega_p = 2\pi \cdot 50$ THz. We assume an $e^{-j\omega t}$ time convention throughout the paper. Using the transverse resonance condition for this simple geometry, the even and odd dispersion relations are found as [76]

$$\tan\left(\frac{k_{1z}t}{2}\right) = j\frac{Z_0}{Z_1} = j\frac{k_{0z}\varepsilon_{enz}}{k_{1z}}, \qquad (2a)$$

$$\cot\left(\frac{k_{1z}t}{2}\right) = j\frac{Z_0}{Z_1} = j\frac{k_{0z}\varepsilon_{enz}}{k_{1z}}. \qquad (2b)$$

The parameters $Z_0$, $k_{0z} = \sqrt{k_0^2 - k_x^2}$ and $Z_1$, $k_{1z} = \sqrt{k_0^2\varepsilon_{enz} - k_x^2}$ are transverse magnetic (TM) wave impedances and propagation constants along *z*-direction in air and ENZ, respectively, $k_0 = \omega/c$ is the wavenumber in free space, c is the speed of light in air and *t* is the slab thickness, and $k_x$ is the transverse wave number, assumed to be real. Figure 1(a) shows the even TM modal dispersion for the slab (blue dots), representing the real part of the mode eigenfrequency obtained through equation (2a). We note that, since we focus on leaky modes, the supported surface plasmon modes are omitted from the discussion and the dispersion diagram.

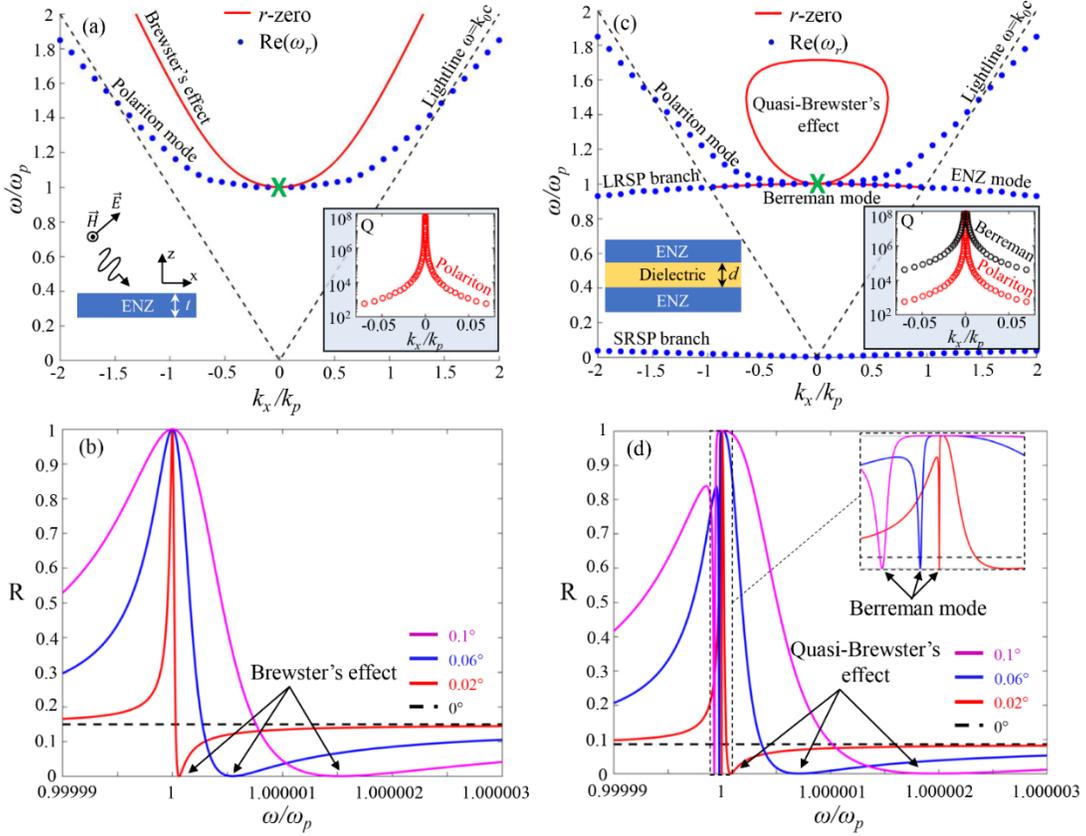

**Figure 1: Modal dispersion and analysis of the reflection coeffcient.** (a) Mode dispersion for an ENZ slab with $t = 800$ nm. Reflection zeros (red line), and real part of the eigenfrequency Re($\omega_r$), where the eigenfrequency is defined by $\omega_r = \omega_{re} + j\omega_{im}$ (blue dots). Inset: $Q$-factor of the polariton mode, calculated as $Q = -\omega_{re}/2\omega_{im}$. (b) Reflection coefficient for TM excitation at small angles for (a). (c) Mode dispersion for the ENZ-dielectric-ENZ structure, $t = 400$ nm, $d = 20$ nm ($d \ll \lambda_{diel}$), $\varepsilon_d = 10$. Inset: $Q$-factor of the polariton and Berreman modes. (d) Reflection coefficient for TM excitation at small angles for (c). Plasma frequency is $f_p = \omega_p/2\pi = 50$ THz. The frequency axis is normalized to $\omega_p$, while the wavenumber axis is normalized to $k_p = \omega_p/c$.

Since the slab thickness is subwavelength in the frequency regime of interest, the only available radiative mode is the bulk plasmon, also called the polariton mode [47]. This mode is leaky as it enters the light cone, and in this region an imaginary part arises even when material loss are not considered. The Q-factor of the polariton mode is therefore finite, accounting for radiation loss, and it is plotted in a range of angles close to normal incidence in the inset of Figure 1(a). It is worth noting that the Q-factor of the modes contains the information on their imaginary frequency as Q=-$\omega_{re}$/2$\omega_{im}$. We observe in particular that the Q-factor diverges at $k_x = 0$, indicating the presence of a symmetry-protected embedded eigenstate, for which the imaginary part of the eigenfrequency goes to zero, consistent with the findings in [50]. Since the structure is lossless, time-reversal symmetry requires that, as the pole approaches the real axis, a zero of the *S*-matrix eigenvalues also meets the pole for the same frequency, and their degeneracy produces an embedded eigenstate [76].

The scattering response of this slab can also be analyzed through the reflection and transmission coefficients, which can be obtained using the ABCD matrix formulation [67], [68], [76] and expressed as

$$r = \frac{j\frac{Z_1^2 - Z_0^2}{2Z_1 Z_0}\sin(k_{1z}t)}{\cos(k_{1z}t) + j\frac{Z_1^2 + Z_0^2}{2Z_1 Z_0}\sin(k_{1z}t)} = 0. \tag{3a}$$

$$t = \frac{1}{\cos(k_{1z}t) + j\frac{Z_1^2 + Z_0^2}{2Z_1 Z_0}\sin(k_{1z}t)} = 0. \tag{3b}$$

where the TM wave impedances are defined as $Z_0 = \eta_0 \cos\theta_i$ and $Z_1 = \eta_0 \cos\theta_1/\sqrt{\varepsilon_{enz}}$, $\eta_0$ is the free-space wave impedance, $\theta_i$ is the incident angle and $\theta_1$ is the angle of refraction in the ENZ layer.

The complex zeros of the denominators in the reflection/transmission coefficients correspond to the solutions of the dispersion relation Eq. (2), i.e., they represent the eigenmodes of the system. It is also interesting to analyze the conditions for which the zeros of reflection (full transmission) occur. For this purpose, we next consider purely real frequencies, as the relevant reflection zeros appear on the real frequency axis for lossless systems. If we assume that $k_{1z}$ is real, i.e., we operate below the critical angle $|\theta_C| = \arcsin\sqrt{\varepsilon_{enz}}$, then the phase advance through the slab is negligible for small thicknesses, $k_{1z}t \ll 1$. Hence, since $\sin(k_{1z}t) \neq 0$, no Fabry-Perot modes occur and the only way to obtain a reflection zero is through non-resonant impedance matching, as pointed out in [74, 75]. In other words, full transmission can occur when the transverse wave impedance of air and ENZ are matched $Z_0 = Z_1$ [expression (3a) and (3b)], which corresponds to the Brewster's condition [76]

$$|\theta_B| = \arctan\left(\frac{\sqrt{\varepsilon_{enz}}}{\sqrt{\varepsilon_{air}}}\right) = \arctan\left(\sqrt{1 - \left(\frac{\omega_p}{\omega}\right)^2}\right), \omega \geq \omega_p. \tag{4}$$

Since the slab has a Drude permittivity dispersion, the Brewster's angle for this problem is frequency-dependent and it exists only for positive values of permittivity, i.e., above the plasma frequency. The tunneling occurs at the Brewster frequency

$$\omega_B = \frac{\omega_p}{\sqrt{1 - \tan^2 \theta_i}}, -\frac{\pi}{4} < \theta_i < \frac{\pi}{4}. \tag{5}$$

This Brewster tunneling effect is illustrated in Figure 1(a) as the zero-reflection line (red line), while Figure 1(b) shows the reflection coefficient dispersion for TM polarized light impinging on the slab at angles close to the normal incidence, where the incident angle is defined as $\theta_i = arcsin(|k_x/k_0|)$. As the incident angle gets closer to zero, the Brewster's condition gets closer to the plasma frequency, expression (5), and to the embedded eigenstate at at $\theta_i = 0°$, with the associated linewidth narrowing. In the limit, the pole and zeros merge on the real axis and an embedded eigenstate arises, marked by the green cross in Figure 1(a). The embedded eigenstate is a dark mode that does not radiate, hence its infinite lifetime, and therefore cannot be observed through external excitation because of reciprocity.

Next, we introduce a subwavelength dielectric gap within the ENZ layer, as in the inset of Figure 1(c). Following the previous steps, we again use the transverse resonance method and ABCD matrix formulation to obtain the resonant condition and reflection coefficient for the 3-layer structure [76]. For the sake of brevity, we provide here only the dispersion relation for even modes:

$$\tan\left(\frac{k_{2z}d}{2}\right) = j \frac{k_{1z}\varepsilon_d}{k_{2z}\varepsilon_{enz}} \frac{\left(k_{0z} + j\frac{k_{1z}}{\varepsilon_{enz}}\tan(k_{1z}t)\right)}{\left(\frac{k_{1z}}{\varepsilon_{enz}} + jk_{0z}\tan(k_{1z}t)\right)}, \tag{6a}$$

and odd modes:

$$\cot\left(\frac{k_{2z}d}{2}\right) = j\frac{k_{1z}\varepsilon_d}{k_{2z}\varepsilon_{enz}}\frac{\left(k_{0z}+j\frac{k_{1z}}{\varepsilon_{enz}}\tan(k_{1z}t)\right)}{\left(\frac{k_{1z}}{\varepsilon_{enz}}+jk_{0z}\tan(k_{1z}t)\right)}. \qquad (6b)$$

The parameters *d = 20 nm (d<<λ_{diel})* and *t* = 400 nm are the thickness of the dielectric layer and ENZ layers, respectively, the relative permittivity of the dielectric slab $\varepsilon_d$ is equal to 10 and $k_{nz}$ is the propagation constant along the *z*-axis in each layer.

Figure 1(c) and 1(d) show the dispersion diagram and reflection coefficient for this geometry. Again, the surface modes irrelevant for the following discussion have been omitted from the dispersion diagram for clarity Above the plasma frequency, we note the same polariton mode as in the single-slab case, as well as the occurrence of a tunneling phenomenon, analogous to Brewster, denoted by the red closed contour, Figure 1(c). Similarly to the single-slab case, the non-resonant impedance matching and consequently the reflection zeros occur above the plasma frequency due to a quasi-Brewster's effect, with the difference here that this tunneling depends on the effective permittivity of the 3-layer structure. The peculiarity of this dispersion diagram is a new mode, emerging slightly below the plasma frequency, called the *Berreman* mode. This phenomenon was first discovered observing the narrow absorption peaks in dielectric-thin film multilayers [57], and it represents a leaky bulk plasmon mode above the light line. The extension of the Berreman mode below the light-line is also known as the ENZ mode [58], which is supported by a long-range surface plasmon (LRSP) branch in the case of a very thin metal [59]. Due to the *manifold* nature of the Berreman mode, the LRSP and ENZ branches are also denoted in the dispersion diagram, although these guided modes are not of particular interest for the phenomena discussed here. In addition, a short-range surface plasmon (SRSP) branch is also found below the light-line for lower frequencies.

The inset of Figure 1(c) shows the Q-factor of both polariton and Berreman modes at $k_x$=0. Their divergence indicates that both modes support symmetry-protected embedded eigenstates, which is also confirmed by the merging of the corresponding poles and zeros of the *S*-matrix eigenvalues at the real frequency axis [76]. Moreover, the Berreman and polariton modes become degenerate at normal incidence, with both eigenfrequencies equal and real, thus showing that this system supports a doubly-degenerate symmetry protected EE. In terms of reflection, Figure 1(d), there are now two zero-reflection contours approaching each other around normal incidence, as opposed to the case of single ENZ slab.

The Berreman mode has mostly been studied in the context of multilayer metal-dielectric structures for various applications [33], [59]-[66]. Its physical nature has been primarily explained through effective medium theory [36],[61], noticing how the spectral position of this resonance corresponds to the frequency at which the effective permittivity of the structure crosses zero. Namely, stacking metal-dielectric pairs can enable a hyperbolic dispersion [63], and induce effective ENZ response using the transition between elliptic and hyperbolic propagation [33]. Another theoretical description of the Berreman mode using a harmonic oscillator model was recently put forward [66], relating tunneling at the Berreman mode in a metal-insulator-metal cavity to electron tunneling through a double-barrier potential well.

To provide physical insights into its role in the context of embedded eigenstates, the Berreman mode is here studied using a simple analogy to a lumped circuit element [76]. Namely, below the plasma frequency and in the limit of extremely subwavelength dielectric thicknesses, this open structure acts like a lossless lumped-element *L-C-L* circuit, loaded by $Z_0$ on both sides to model radiation. The permittivity of the ENZ slab is negative below the plasma frequency, making its wave impedance inductive, as well as making the wave propagation in the ENZ purely evanescent, i.e., without phase advance. On the other hand, the subwavelength dielectric gap has a capacitive impedance, with a negligible phase advance. Thus the resonant behaviour of the multilayer is analogous to a lumped element *L-C-L* circuit. The tunneling feature of the Berreman mode can then be attributed to a classical *LC* resonance, i.e., resonant impedance matching, where the reactance contributions of the inductor (ENZ slab) and capacitor (dielectric gap) cancel out, balancing the magnetic and electric energy of the oscillator circuit. To distinguish this mode from others in the following discussion, we designate it as the $0^{th}$ order Berreman mode.

To further investigate the potential of the 3-layer structure, we increase the gap size such that its thickness is equal to half of the longitudinal wavelength in the material at an angle of 50°. The corresponding dispersion diagram is shown in Figure 2(a). In this case the $0^{th}$ order Berreman mode shifts to lower frequencies, and its flat dispersion indicates its slow light nature, which can be used for enhanced light-matter interactions [36],[64]-[66]. The angular bandwidth of this slow-light regime is exceptionally large, spanning across most of the light cone. Below the light line, the LRSP and SRSP modes occur, with the well-known asymptote $\omega = \omega_p/\sqrt{1+\varepsilon_d}$.

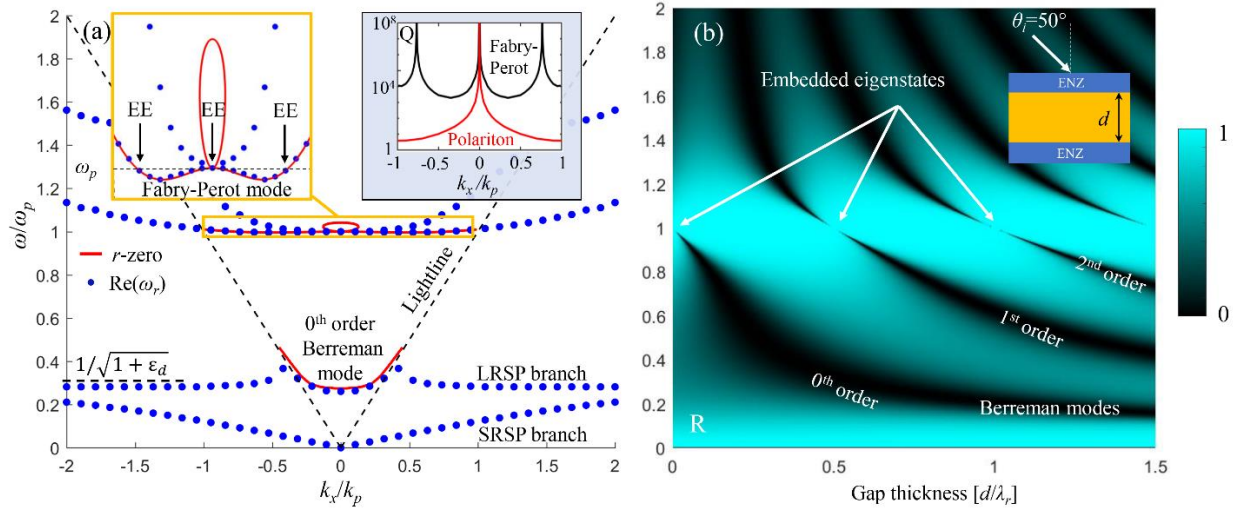

**Figure 2: Higher-order Berreman modes and accidental EEs.** (a) Mode dispersion for a half-wavelength thick dielectric gap, supporting both normal and off-normal incidence ($\theta_i = 50$ deg) embedded eigenstates, $d = 980$ nm, $\varepsilon_d = 10$. Inset: zoom-in around the plasma frequency (left), and Q-factor of the modes (right). (b) Reflection coefficient for TM polarized wave at $\theta_i=50$ deg for different dielectric thicknesses, where different orders of Berreman modes and accidental EEs are visible.

Since the dielectric is half-wavelength thick, it also supports a Fabry-Perot mode. The precise resonant thickness for which a non-zero longitudinal wavenumber $k_x$ supports such a resonance (corresponding to an incidence angle $\sin(\theta_i) = |k_x/k_o|$) is given by

$$d = \frac{\lambda_r}{2} = \frac{c}{2f_r\sqrt{\varepsilon_d - \left(\frac{k_x}{k_0}\right)^2}} = \frac{c}{2f_r\sqrt{\varepsilon_d - \sin^2\theta_i}} \quad (7)$$

where $\lambda_r$ is the resonant wavelength, $f_r$ is the resonant frequency, $c$ is the speed of light in vacuum. If the dispersion of this mode crosses the plasma frequency for some angle, i.e., if $f_r=f_p$, an off-normal (accidental) EE forms, for which the mode is perfectly confined without radiation leakage [27]. This Fabry-Perot mode, which represents a 1st order Berreman mode, supports full energy tunneling. The described system now supports two embedded eigenstates: the symmetry-protected one at normal incidence and an accidental one. Figure 2(a, inset) shows the dispersion around the plasma frequency, where these EEs are visible.

In order to better describe the effect of the dielectric thickness on the modal evolution, we plot the reflection coefficient for TM polarized waves at an incidence angle of 50 deg for different dielectric thicknesses $d$, Figure 2(b). The vanishing linewidth can be observed at the plasma frequency for $d = n\lambda_r/2$, corresponding to an embedded eigenstate of the $n^{th}$ order Berreman mode. It is worth noting that for $n = 0$ the system corresponds to a single ENZ slab, which makes this EE a trivial one.

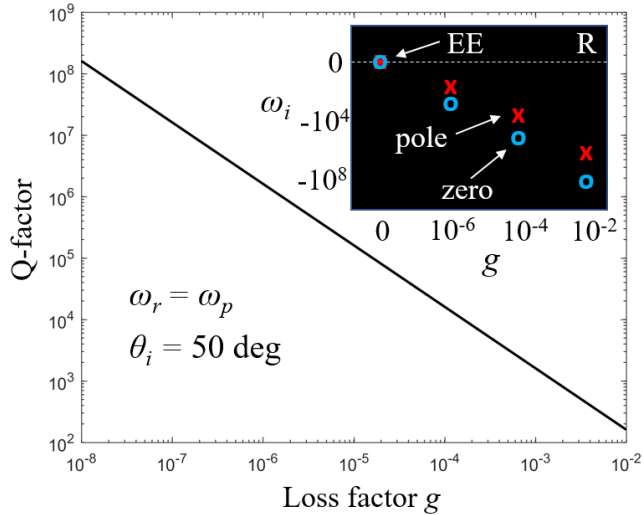

**Figure 3: Effect of loss on the accidental embedded eigenstate**. $Q$-factor as a function of loss factor $g$. Inset shows a sketch of the position of complex poles and zeros of reflection as a function of $g$.

**Loss in ENZ.** The role of loss in ENZ-based devices cannot be understated, as they are unavoidable in realistic materials and represent a limiting factor in many applications [69]. Any material carries intrinsic loss, and when they are paired with the low group velocity of ENZ materials, even small loss can become critical. To study the effect of losses, we plot the Q-factor of the localized Fabry-Perot mode as a function of the loss factor $g$, Figure 3. The loss of the ENZ material has been incorporated in the Drude model using $\varepsilon_{enz} = 1 - \omega_p^2/(\omega^2 + i\gamma\omega)$, where $\gamma = g\omega_p$ ($g = \text{Im}[\varepsilon_{enz}(\omega_p)]$). We fix the incidence angle again to 50 deg and the frequency to $\omega_p$, at which the system supports an embedded eigenstate, and calculate the reflection coefficient in the complex frequency space as a function of $g$. The reflection poles correspond to the eigenmodes of the system with complex frequency $\omega_r = \omega_{re} + j\omega_{im}$, and therefore the Q-factors can be readily calculated as $\omega_r/2\omega_i$. It can be seen that there is a linear relation between $g$ and Q-factor, with $Q \sim 1/g$, putting a rather restrictive limit on the achievable Q-factors.

Another critical effect associated with the material loss can be highlighted by analyzing the poles and zeros of reflection. The Fabry-Perot resonance in this system is a 1[st] order Berreman mode, supporting complete tunneling of energy in the lossless limit, with all of the zeros of reflection pinned to the real frequency axis (zero imaginary part). This tunneling persists along the whole Fabry-Perot dispersion line, except at the EE where the reflection has both a pole and a zero at the same real frequency, i.e., the reflection coefficient is undefined in the lossless limit. By adding loss, the pole and zero split and move down in the complex plane, making this state a quasi-EE (qEE). Interestingly, the pole stays *closer* to the real frequency axis than the zero, as shown in Figure 3 (inset). This peculiarity has important implications: in the lossy case, the tunneling of the Berreman mode vanishes at the angle corresponding to the quasi-EE. In parallel, the reflection coefficient becomes larger than zero and approaches unity at the qEE. Indeed, the transition from metal to dielectric at the plasma frequency necessarily results in strong reflectivity of the ENZ layers and significant impedance mismatch with free space, which persist against added losses. As a consequence, the resonant line vanishing in the realistic lossy system happens not only due to the $Q$-factor enhancement but also due to inevitable reflection at the qEE.

**Realistic structure comprising SiC.** Although the discussed effects of material loss may fundamentally limit the $Q$-factor of the quasi-EEs supported by the proposed geometry, low-loss ENZ materials can nevertheless be used to achieve high-Q quasi-EE resonances, of interest for selective transmission and sensing. Here, we explore a realistic structure with silicon carbide as the ENZ material that supports high-Q quasi-EE at 50 deg incidence, Figure 4. SiC is an excellent candidate for this purpose since it displays very low loss at the longitudinal phonon frequency where the real part of permittivity crosses zero [45], [70], [71]. The permittivity of 4H-SiC can be approximated as[45],[71]

$$\varepsilon_{\text{SiC}} = \varepsilon_\infty \left( 1 + \frac{(\omega_{LO}^2 - \omega_{TO}^2)}{\omega_{TO}^2 - \omega^2 + i\gamma\omega} \right), \tag{8}$$

where $\omega_{LO}=2\pi \cdot 29.08$ THz and $\omega_{TO}=2\pi \cdot 23.89$ THz are longitudinal and transverse optical phonon frequencies, $\gamma=2\pi \cdot 0.04$ THz is damping, $\varepsilon_\infty=6.6$.

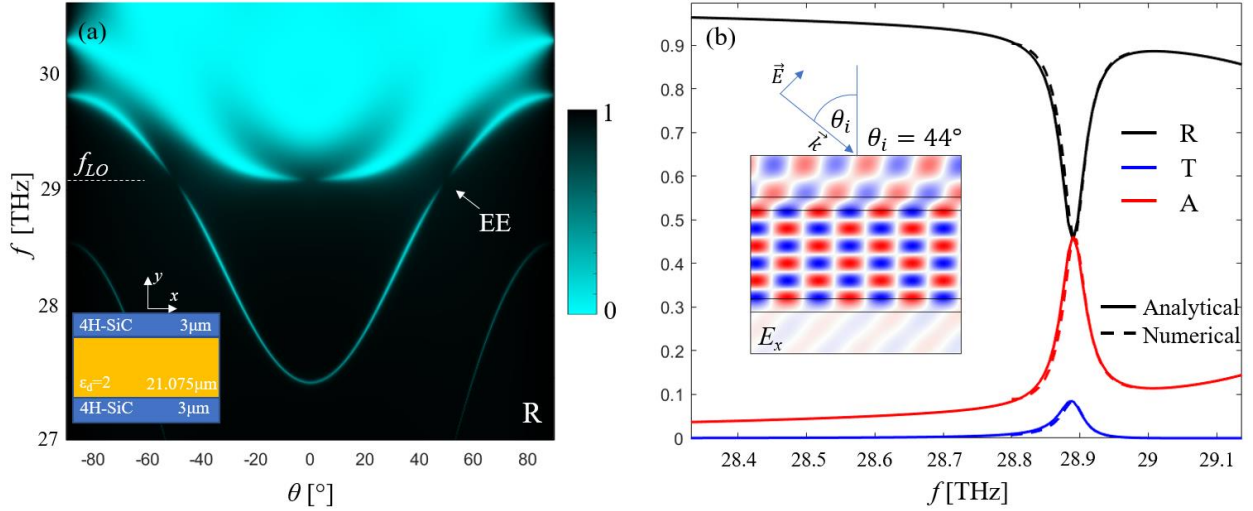

**Figure 4**: **Demonstration of quasi-EE in SiC.** (a) Reflection spectrum for TM-polarized light with the geometry sketched in the inset. (b) Reflection, transmission, and absorption of a higher-order Berreman mode at 44 deg incidence. Inset: E-field distribution inside the structure.

Figure 4(a) shows the reflection spectrum of a SiC-dielectric-SiC cavity, displaying a q-EE at the angle of incidence of 50 deg. The high-Q Berreman mode supported by this structure has a dispersion crossing the $\omega_{LO}$ frequency of SiC where the ENZ regime arises. This is corroborated by vanishing of the resonant line in the reflection spectrum. A numerical demonstration of a resonance close to the EE point is pictured in Figure 4(b), with Q ~$10^3$, and the scattering coefficients are in excellent agreement with the analytical ones obtained with the ABCD matrix approach [76] In case we use a more conservative estimation of loss Im($\varepsilon$)= 0.07 as in [51], we obtain Q factors ∼ 320.

**Quasi-EE for narrowband perfect absorption and thermal emission.** The ENZ regime implies strong light-matter interactions with ultra-narrow scattering lines. In the case of a realistic lossy ENZ, this property results in enhanced absorption near the ENZ point, as shown in Figure 4(b), and this property can be used to engineer narrowband absorbers based on quasi-embedded eigenstates [51]. Following the conventional approach to obtain perfect absorption [72], we reduce transmission through the structure by increasing the thickness of the lower SiC layer and consider the asymmetric structure geometry in Figure 5(a). As discussed above, incorporation of loss ensures a reflection maxima exactly at the q-EE, which results in weak absorption. However, the resonances around the q-EE can display absorption peaks with extremely narrow angular and frequency bandwidth.

To exploit the potential of the proposed multilayer structure to its full extent, we first analyse the absorption of the single-slab and the asymmetric 3-layer structures, Figures 5(a) and (d). Both the polariton mode in a thin free-standing ENZ layer, Figure 5(d), and the higher-order Berreman (Fabry-Perot) mode in the 3-layer structure, Figure 5(a), have absorption features that are strongly related to the spectral position of the ENZ point. However, the polariton mode does not allow efficient dispersion control, has poor angular selectivity and does not support engineering of perfect absorption [76]. Although the dispersion of the Fabry-Perot mode in the 3-layer structure can be controlled with the resonator itself, thus potentially improving the angular and frequency bandwidth of the absorption peaks, the structure in Figure 5(a) has poor angular selectivity, since the dispersion of the Fabry-Perot mode follows the ENZ frequency $\omega_{LO}$ closely for all incident angles where the absorption is enhanced.

To change the dispersion of the mode and further improve selectivity, we incorporate a distributed Bragg reflector (DBR) between the resonator and each SiC layer, Figure 5(b,c). While the proposed structure can get quite bulky, the spectral and angular bandwidths become exceptionally narrow for the long-wavelength IR region. The introduction of a DBR expectedly narrows the spectral width of the resonance, yet it introduces a steep angular dispersion of the Fabry-Perot mode as well, moving it further away from the ENZ frequency. This, in turn, enables a narrower angular width of near-perfect absorption regions around the quasi-EE. These features abruptly disappear when one of the SiC is removed, as shown in Figures 5(e,f), which demonstrates the importance of the ENZ-resonator-ENZ configuration to yield this exotic absorption and thermal emission features. Figure 5(b) shows near-perfect absorption at two points near the q-EE, which are both spectrally and spatially narrowband. Both the absorption peak and the large Q-factor are a direct result of the q-EE in the ENZ regime, indicating its pivotal role in engineering extreme light-matter interactions and absorption with extremely narrow spectral and angular features.

Given the extreme spectral and spatial selectivity of the proposed absorbing structure, and the fact that operation frequency falls within the thermal long-wave infrared window (8-12μm), it represents an ideal candidate for the design of quasi-coherent thermal emitters, given Kirchhoff's law that directly relates optical absorption and thermal emission [73].

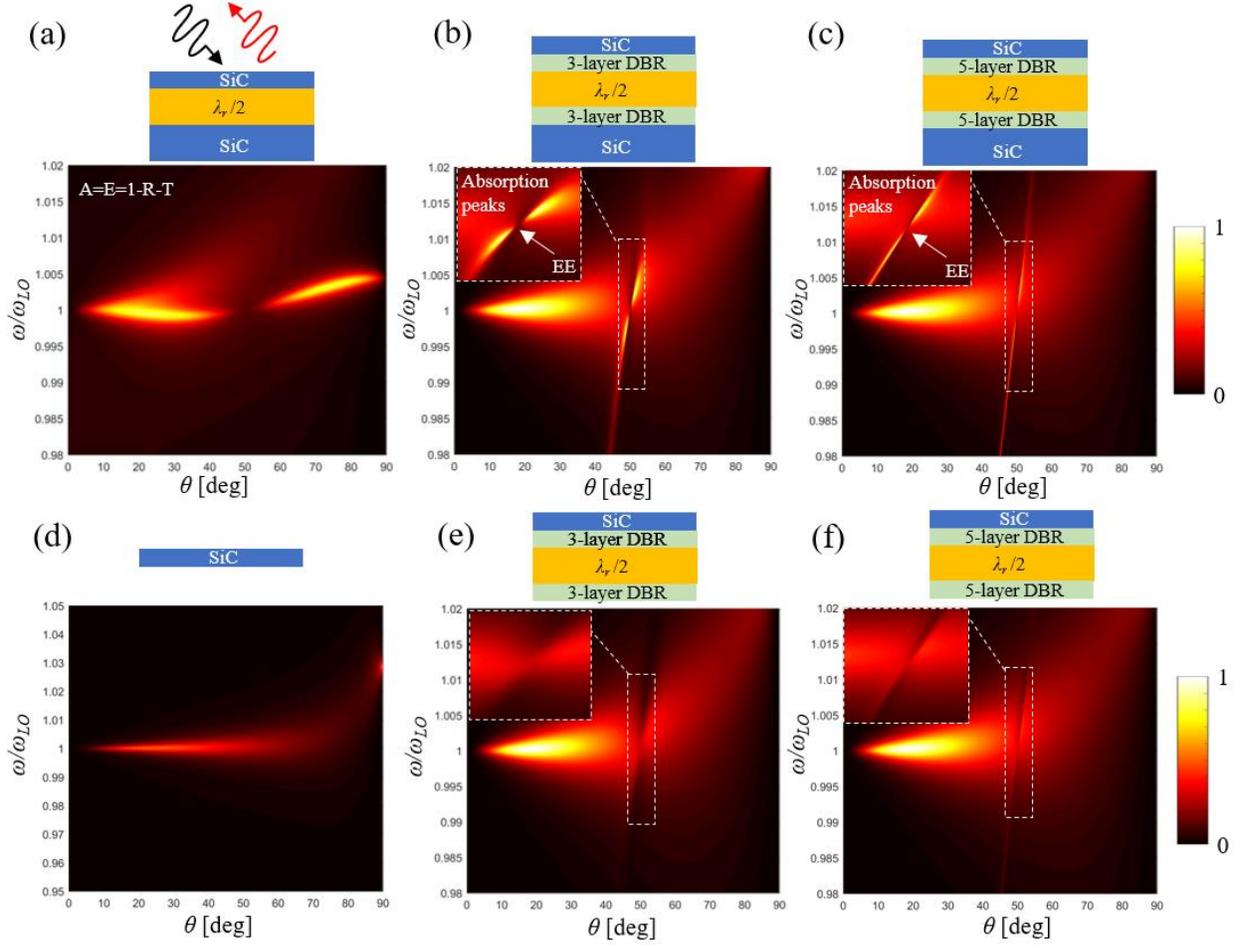

**Figure 5: Absorption for TM illumination from the top with schematics of the multilayer structures under investigation.** (a) Asymmetric SiC-Dielectric-SiC structure supporting near-perfect absorption of the Fabry-Perot mode. (b) Absorption spectrum for a multilayer structure with 3-layer DBR, and close-up view around q-EE. (c) Absorption spectrum for a multilayer structure with 5-layer DBR, and close-up view around q-EE, (d) Single SiC slab, (e) same as (b) with removed bottom SiC layer, (f) same as (c) with removed bottom SiC layer. Top SiC layer thickness is $t_T$ =500 nm, bottom is $t_B$ = 1500 nm. High and low permittivity quarter-wave thickness' are $d_H$ = 656 nm and $d_L$ = 2.168 μm, while the resonator thickness is $d_R$ = 4.33 μm. Resonator and the low permittivity slab of the DBR are made of a low-loss low permittivity material, modeled after BaF$_2$ with $\varepsilon_r \approx 2$ around 10μm. The high permittivity slab for DBR used here is made of Ge which has $\varepsilon_r \approx 16$ and negligible losses around 10μm.

To demonstrate the potential of the discussed configuration, we study a multilayer structure with 5-layer DBR, as pictured in the inset of Figure 6(a). The absorption (emission) lines for angles around the embedded eigenstate display a very narrow band, with minimum absorption exactly at the q-EE as discussed previously. To compare the thermal performance of the structure with a single slab of SiC, we plot spectral radiance in the region 8-12 μm for the same structure, Figure 6(b). Emission peak near the qEE demonstrates extreme spectral selectivity, with a linewidth of less than 10nm at half-maximum, and consequently superiority in comparison to the single-layer SiC structure. With an increasing number of layers in the DBR, the linewidth narrows further, but also the structure gets bulkier.

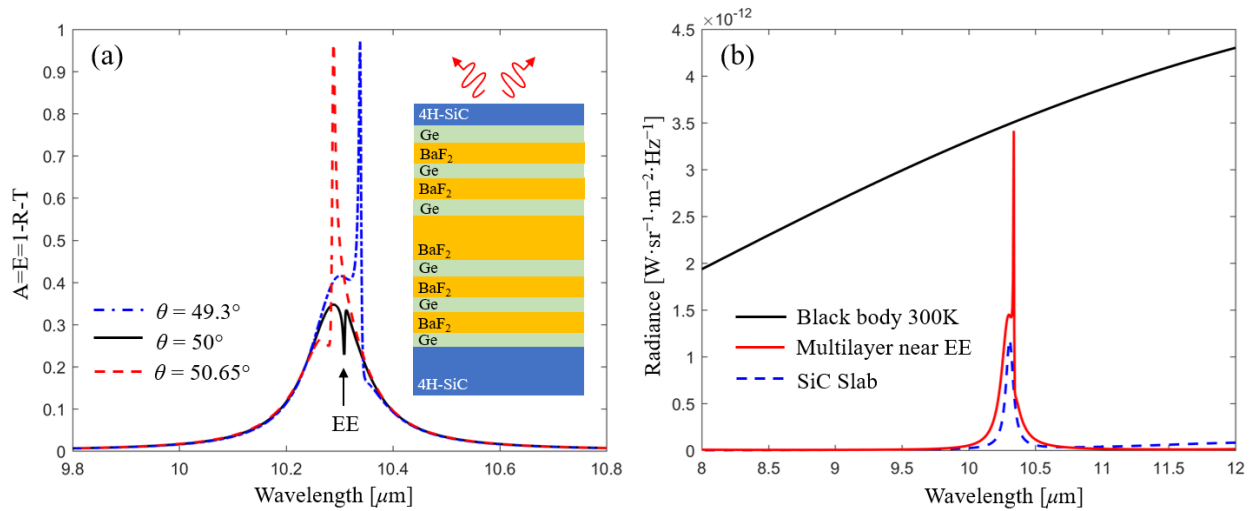

**Figure 6: Emission near EE:** Multilayer structure, with the same parameters as in figure 5 except for optimized $t_t$=400 nm and $t_B$=2 μm, with 19.35μm thickness in total. (a) Absorption/thermal emission around EE (b) TM spectral radiance of the proposed structure at 49.3° compared to a single SiC slab and black body radiance at 300K.

## Conclusions

We have presented an in-depth investigation of modes supported by planar structures comprising ENZ materials, identifying several key features of these modes. Specifically, we have explained the origin and nature of leaky Berreman modes and how they relate to embedded eigenstates using a simple model based on the transverse resonance technique. We have identified two types of EEs in such structures – symmetry-protected EEs at normal incidence and accidental EEs at a desired angle of incidence. For practical considerations, we have proposed a realizable SiC-based structure that supports quasi-EEs with high Q-factors and field enhancements. Based on these concepts, we have demonstrated extremely narrowband absorber/thermal emitters near the q-EE. In light of recent findings [77], we believe that the incorporation of optical nonlocality in ENZ materials can possibly provide a new degree of freedom in engineering EEs and related scattering effects. This is of great interest for thermal engineering, as the understading of nonlocality in polar dielectric grows [78]. We believe that the presented concepts can offer exciting opportunities for ENZ-based devices in the context of thermal engineering, sensing, non-linear optics, and filtering.

## Acknowledgements

The work described in this paper has been conducted within the project NOCTURNO that has received funding from the European Union's Horizon 2020 research and innovation programme under Grant No. 777714. This work has been also partially supported by the Department of Defense Vannevar Bush Faculty Fellowship, the Simons Foundation and the National Science Foundation.

## References

[1]   K. J. Vahala, "Optical microcavities," *Nature*, vol. 424, no. 6950, pp. 839–846, 2003.

[2]   V. S. Ilchenko and A. B. Matsko, "Optical resonators with whispering-gallery modes - Part II: Applications," *IEEE J. Sel. Top. Quantum Electron.*, vol. 12, pp. 15–32, 2006.

[3]   A. Krasnok, D. Baranov, H. Li, M.-A. Miri, F. Monticone, and A. Alú, "Anomalies in light scattering," *Adv. Opt. Photonics*, vol. 11, no. 4, p. 892, Dec. 2019.

[4]   K. Koshelev, S. Lepeshov, M. Liu, A. Bogdanov, and Y. Kivshar, "Asymmetric Metasurfaces with High-Q Resonances Governed by Bound States in the Continuum," *Phys. Rev. Lett.*, vol. 121, no. 19, p. 193903, Nov. 2018.

[5]   K. Fan, I. V. Shadrivov, and W. J. Padilla, "Dynamic bound states in the continuum," *Optica*, vol. 6, no. 2, p. 169, Feb. 2019.

[6]   J. von Neumann and E. P. Wigner, "Über merkwürdige diskrete Eigenwerte," in *The Collected Works of Eugene Paul Wigner*, Berlin, Heidelberg: Springer Berlin Heidelberg, 1993, pp. 291–293.

[7]   D. C. Marinica, A. G. Borisov, and S. V. Shabanov, "Bound States in the Continuum in Photonics," *Phys. Rev. Lett.*, vol. 100, no. 18, p. 183902, May 2008.

[8]   C. W. Hsu, B. Zhen, A. D. Stone, J. D. Joannopoulos, and M. Soljačić, "Bound states in the

continuum," *Nat. Rev. Mater.*, vol. 1, no. 9, p. 16048, Sep. 2016.

[9] A. Kodigala, T. Lepetit, Q. Gu, B. Bahari, Y. Fainman, and B. Kanté, "Lasing action from photonic bound states in continuum," *Nature*, vol. 541, no. 7636, pp. 196–199, Jan. 2017.

[10] T. Lepetit and B. Kanté, "Controlling multipolar radiation with symmetries for electromagnetic bound states in the continuum," *Phys. Rev. B - Condens. Matter Mater. Phys.*, vol. 90, no. 24, pp. 1–4, 2014.

[11] E. N. Bulgakov and A. F. Sadreev, "Bloch bound states in the radiation continuum in a periodic array of dielectric rods," *Phys. Rev. A - At. Mol. Opt. Phys.*, vol. 90, no. 5, pp. 1–7, 2014.

[12] M. V. Rybin *et al.*, "High- Q Supercavity Modes in Subwavelength Dielectric Resonators," *Phys. Rev. Lett.*, vol. 119, no. 24, p. 243901, Dec. 2017.

[13] S. I. Azzam, V. M. Shalaev, A. Boltasseva, and A. V. Kildishev, "Formation of Bound States in the Continuum in Hybrid Plasmonic-Photonic Systems," *Phys. Rev. Lett.*, vol. 121, no. 25, p. 253901, 2018.

[14] E. A. Bezus, D. A. Bykov, and L. L. Doskolovich, "Bound states in the continuum and high-Q resonances supported by a dielectric ridge on a slab waveguide," *Photonics Res.*, vol. 6, no. 11, p. 1084, Nov. 2018.

[15] H. M. Doeleman, F. Monticone, W. Den Hollander, A. Alù, and A. F. Koenderink, "Experimental observation of a polarization vortex at an optical bound state in the continuum," *Nat. Photonics*, vol. 12, no. 7, pp. 397–401, Jul. 2018.

[16] A. Krasnok and A. Alu, "Embedded scattering eigenstates using resonant metasurfaces," *J. Opt. (United Kingdom)*, vol. 20, no. 6, 2018.

[17] C. W. Hsu *et al.*, "Observation of trapped light within the radiation continuum," *Nature*, vol. 499, no. 7457, pp. 188–191, 2013.

[18] Y. Plotnik *et al.*, "Experimental observation of optical bound states in the continuum," *Phys. Rev. Lett.*, vol. 107, no. 18, pp. 28–31, 2011.

[19] S. Lannebère and M. G. Silveirinha, "Optical meta-atom for localization of light with quantized energy," *Nat. Commun.*, vol. 6, no. 1, p. 8766, Dec. 2015.

[20] S. D. Krasikov, A. A. Bogdanov, and I. V. Iorsh, "Nonlinear Bound States in the Continuum in One-Dimensional Photonic Crystal Slab," *J. Phys. Conf. Ser.*, vol. 1092, no. March, p. 012068, Sep. 2018.

[21] K. Koshelev, Y. Tang, K. Li, D. Choi, G. Li, and Y. Kivshar, "Nonlinear Metasurfaces Governed by Bound States in the Continuum," *ACS Photonics*, vol. 6, no. 7, pp. 1639–1644, Jul. 2019.

[22] A. A. Bogdanov *et al.*, "Bound states in the continuum and Fano resonances in the strong mode coupling regime," *Adv. Photonics*, vol. 1, no. 01, p. 1, Jan. 2019.

[23] L. S. Li and H. Yin, "Bound states in the continuum in double layer structures," *Sci. Rep.*, vol. 6, no. May, p. 26988, Jul. 2016.

[24] M. Zhao and K. Fang, "Mechanical bound states in the continuum for macroscopic optomechanics," *Opt. Express*, vol. 27, no. 7, p. 10138, Apr. 2019.

[25] A. A. Lyapina, D. N. Maksimov, A. S. Pilipchuk, and A. F. Sadreev, "Bound states in the continuum in open acoustic resonators," *J. Fluid Mech.*, vol. 780, no. 2015, pp. 370–387, 2015.

[26] S. Han *et al.*, "All-Dielectric Active Terahertz Photonics Driven by Bound States in the Continuum," *Adv. Mater.*, no. 1803.01992, p. 1901921, 2019.

[27] F. Monticone, H. M. Doeleman, W. Den Hollander, A. F. Koenderink, and A. Alù, "Trapping Light in Plain Sight: Embedded Photonic Eigenstates in Zero-Index Metamaterials," *Laser Photon. Rev.*, vol. 12, no. 5, p. 1700220, May 2018.

[28] M. G. Silveirinha, "Trapping light in open plasmonic nanostructures," *Phys. Rev. A - At. Mol. Opt. Phys.*, vol. 89, no. 2, pp. 1–10, 2014.

[29] F. Monticone and A. Alù, "Embedded Photonic Eigenvalues in 3D Nanostructures," *Phys. Rev. Lett.*, vol. 112, no. 21, p. 213903, May 2014.

[30] I. Liberal and N. Engheta, "Nonradiating and radiating modes excited by quantum emitters in open epsilon-near-zero cavities," *Sci. Adv.*, vol. 2, no. 10, pp. e1600987–e1600987, Oct. 2015.

[31] B. Edwards, A. Alù, M. E. Young, M. Silveirinha, and N. Engheta, "Experimental verification of epsilon-near-zero metamaterial coupling and energy squeezing using a microwave waveguide," *Phys. Rev. Lett.*, vol. 100, no. 3, pp. 1–4, 2008.

[32] I. Liberal, A. M. Mahmoud, Y. Li, B. Edwards, and N. Engheta, "Photonic doping of epsilon-near-zero media," *Science (80-. ).*, vol. 355, no. 6329, pp. 1058–1062, 2017.

[33] R. Maas, J. Parsons, N. Engheta, and A. Polman, "Experimental realization of an epsilon-near-zero metamaterial at visible wavelengths," *Nat. Photonics*, vol. 7, no. 11, pp. 907–912, Nov. 2013.

[34] S. Feng and K. Halterman, "Coherent perfect absorption in epsilon-near-zero metamaterials," *Phys. Rev. B - Condens. Matter Mater. Phys.*, vol. 86, no. 16, pp. 10–15, 2012.

[35] S. Jahani, H. Zhao, and Z. Jacob, "Switching Purcell effect with nonlinear epsilon-near-zero media," *Appl. Phys. Lett.*, vol. 113, no. 2, p. 021103, Jul. 2018.

[36]  V. Caligiuri, M. Palei, M. Imran, L. Manna, and R. Krahne, "Planar Double-Epsilon-Near-Zero Cavities for Spontaneous Emission and Purcell Effect Enhancement," *ACS Photonics*, vol. 5, no. 6, pp. 2287–2294, 2018.

[37]  A. Anopchenko, L. Tao, C. Arndt, and H. W. H. Lee, "Field-Effect Tunable and Broadband Epsilon-Near-Zero Perfect Absorbers with Deep Subwavelength Thickness," *ACS Photonics*, vol. 5, no. 7, pp. 2631–2637, 2018.

[38]  Y. Li and C. Argyropoulos, "Tunable nonlinear coherent perfect absorption with epsilon-near-zero plasmonic waveguides," vol. 43, no. 8, pp. 1806–1809, 2018.

[39]  S. Vassant *et al.*, "Epsilon-Near-Zero Mode for Active Optoelectronic Devices," vol. 237401, no. DECEMBER, pp. 1–5, 2012.

[40]  M. Z. Alam, S. A. Schulz, J. Upham, I. De Leon, and R. W. Boyd, "Large optical nonlinearity of nanoantennas coupled to an epsilon-near-zero material," *Nat. Photonics*, vol. 12, no. 2, pp. 79–83, Feb. 2018.

[41]  X. Wen *et al.*, "Doubly Enhanced Second Harmonic Generation through Structural and Epsilon-near-Zero Resonances in TiN Nanostructures," *ACS Photonics*, vol. 5, no. 6, pp. 2087–2093, 2018.

[42]  D. de Ceglia, S. Campione, M. A. Vincenti, F. Capolino, and M. Scalora, "Low-damping epsilon-near-zero slabs: Nonlinear and nonlocal optical properties," *Phys. Rev. B*, vol. 87, no. 15, p. 155140, Apr. 2013.

[43]  C. Argyropoulos, P. Y. Chen, G. D'Aguanno, N. Engheta, and A. Alù, "Boosting Optical Nonlinearities in Epsilon-Near-Zero Plasmonic Channels," Physical Review B, Vol. 85, No. 4, 045129 (5 pages), January 27, 2012.

[44]  C. Argyropoulos, G. D'Aguanno, and A. Alù, "Giant Second Harmonic Generation Efficiency and Ideal Phase Matching with a Double Epsilon-Near-Zero Cross-Slit Metamaterial," Physical Review B, Vol. 89, No. 23, 235401 (6 pages), June 3, 2014.

[45]  J. Kim *et al.*, "Role of epsilon-near-zero substrates in the optical response of plasmonic antennas," *Optica*, vol. 3, no. 3, p. 339, Mar. 2016.

[46]  R. Fleury, and A. Alù, "Enhanced Superradiance in Epsilon-Near-Zero Plasmonic Channels," Physical Review B, Rapid Communications, Vol. 87, No. 20, 201101(R) (5 pages), May 10, 2013.

[47]  A. Alù, M. G. Silveirinha, A. Salandrino, and N. Engheta, "Epsilon-near-zero metamaterials and electromagnetic sources: Tailoring the radiation phase pattern," *Phys. Rev. B - Condens. Matter Mater. Phys.*, vol. 75, no. 15, p. 155410, Apr. 2007.

[48]  M. Silveirinha and N. Engheta, "Tunneling of electromagnetic energy through subwavelength channels and bends using ε-near-zero materials," *Phys. Rev. Lett.*, vol. 97, no. 15, p. 157403, Oct. 2006.

[49] I. Liberal and N. Engheta, "Near-zero refractive index photonics," *Nat. Photonics*, vol. 11, no. 3, pp. 149–158, Mar. 2017.

[50] L. Li, J. Zhang, C. Wang, N. Zheng, and H. Yin, "Optical bound states in the continuum in a single slab with zero refractive index," vol. 013801, pp. 1–5, 2017.

[51] R. Duggan, Y. Ra'di, and A. Alù, "Temporally and Spatially Coherent Emission from Thermal Embedded Eigenstates," *ACS Photonics*, vol. 6, no. 11, pp. 2949–2956, Nov. 2019.

[52] F. Monticone and A. Alu, "Leaky-Wave Theory, Techniques, and Applications: From Microwaves to Visible Frequencies," *Proc. IEEE*, vol. 103, no. 5, pp. 793–821, May 2015.

[53] N. Marcuvitz, "On field representations in terms of leaky modes or Eigenmodes," *IRE Trans. Antennas Propag.*, vol. 4, no. 3, pp. 192–194, Jul. 1956.

[54] L. Ge, Y. D. Chong, and A. D. Stone, "Steady-state ab initio laser theory: Generalizations and analytic results," *Phys. Rev. A - At. Mol. Opt. Phys.*, vol. 82, no. 6, p. 063824, Dec. 2010.

[55] D. G. Baranov, A. Krasnok, and A. Alù, "Coherent virtual absorption based on complex zero excitation for ideal light capturing," *Optica*, vol. 4, no. 12, p. 1457, 2017.

[56] V. Grigoriev *et al.*, "Optimization of resonant effects in nanostructures via Weierstrass factorization," *Phys. Rev. A*, vol. 88, no. 1, p. 011803, Jul. 2013.

[57] D. W. Berreman, "Infrared absorption at longitudinal optic frequency in cubic crystal films," *Phys. Rev.*, vol. 130, no. 6, pp. 2193–2198, Jun. 1963.

[58] S. Vassant, J. Hugonin, F. Marquier, and J. Greffet, "Berreman mode and epsilon near zero mode," vol. 20, no. 21, pp. 3995–3998, 2012.

[59] S. Campione, I. Brener, and F. Marquier, "Theory of epsilon-near-zero modes in ultrathin films," *Phys. Rev. B - Condens. Matter Mater. Phys.*, vol. 91, no. 12, pp. 1–5, 2015.

[60] Y. Chen and F. Chiu, "Trapping mid-infrared rays in a lossy film with the Berreman mode, epsilon near zero mode, and magnetic polaritons", Optics Express, vol. 21, no. 18, p. 20771, 2013. Available: 10.1364/oe.21.020771.

[61] W. D. Newman, C. L. Cortes, J. Atkinson, S. Pramanik, R. G. Decorby, and Z. Jacob, "Ferrell-berreman modes in plasmonic epsilon-near-zero media," *ACS Photonics*, vol. 2, no. 1, pp. 2–7, Jan. 2015.

[62] T. Taliercio, V. N. Guilengui, L. Cerutti, E. Tournié, and J. Greffet, "Brewster ' mode ' in highly doped semiconductor layers : an all-optical technique to monitor doping concentration," vol. 22, no. 20, pp. 24294–24303, 2014.

[63] V. V. P. V. Drachev, V. V. A. Podolskiy, and A. A. V. A. Kildishev, "Hyperbolic metamaterials: new physics behind a classical problem," *Opt. Express*, vol. 21, no. 12, pp.

1699–1701, 2013.

[64] M. Y. Shalaginov *et al.*, "Enhancement of single-photon emission from nitrogen-vacancy centers with TiN/(Al,Sc)N hyperbolic metamaterial," *Laser Photonics Rev.*, vol. 9, no. 1, pp. 120–127, 2015.

[65] L. Li, W. Wang, T. S. Luk, X. Yang, and J. Gao, "Enhanced Quantum Dot Spontaneous Emission with Multilayer Metamaterial Nanostructures," *ACS Photonics*, vol. 4, no. 3, pp. 501–508, 2017.

[66] V. Caligiuri, M. Palei, G. Biffi, S. Artyukhin, and R. Krahne, "A Semi-Classical View on Epsilon-Near-Zero Resonant Tunneling Modes in Metal/Insulator/Metal Nanocavities," *Nano Lett.*, vol. 19, no. 5, pp. 3151–3160, 2019.

[67] D. M. Pozar, *Microwave engineering, 4th Edition*. John Wiley & Sons, Inc., 2011.

[68] Y. Li and C. Argyropoulos, "Exceptional points and spectral singularities in active epsilon-near-zero plasmonic waveguides," *Phys. Rev. B*, vol. 99, no. 7, pp. 1–43, 2019.

[69] M. H. Javani and M. I. Stockman, "Real and Imaginary Properties of Epsilon-Near-Zero Materials," *Phys. Rev. Lett.*, vol. 117, no. 10, p. 107404, Sep. 2016.

[70] N. C. Passler *et al.*, "Strong Coupling of Epsilon-Near-Zero Phonon Polaritons in Polar Dielectric Heterostructures," *Nano Lett.*, vol. 18, no. 7, pp. 4285–4292, 2018.

[71] A. Paarmann, I. Razdolski, S. Gewinner, W. Schöllkopf, and M. Wolf, "Effects of crystal anisotropy on optical phonon resonances in midinfrared second harmonic response of SiC," *Phys. Rev. B*, vol. 94, no. 13, pp. 1–9, 2016.

[72] Y. Ra'di, C. R. Simovski, and S. A. Tretyakov, "Thin Perfect Absorbers for Electromagnetic Waves: Theory, Design, and Realizations," *Phys. Rev. Appl.*, vol. 3, no. 3, p. 037001, Mar. 2015.

[73] D. G. Baranov, Y. Xiao, I. A. Nechepurenko, A. Krasnok, A. Alù, and M. A. Kats, "Nanophotonic engineering of far-field thermal emitters," *Nat. Mater.*, vol. 18, no. 9, pp. 920–930, Sep. 2019.

[74] A. Alù, G. D'Aguanno, N. Mattiucci and M. Bloemer. "Plasmonic Brewster Angle: Broadband Extraordinary Transmission through Optical Gratings," *Phys. Rev. Lett.*, Vol. 106, no. 12, 2011.

[75] V. Popov, S. Tretyakov A. Novitsky. (2019). "Brewster effect when approaching exceptional points of degeneracy: Epsilon-near-zero behavior," *Phys. Rev. B*, Vol. 99, no. 4, 2019.

[76] Supplementary materials.

[77] S. Silva, T. Morgado, and M. Silveirinha. "Multiple embedded eigenstates in nonlocal
22

plasmonic nanostructures," *Phys. Rev. B*, Vol. 101, (2020).

[78] C. Gubbin, S. De Liberato, "Optical Nonlocality in Polar Dielectrics," *Phys. Rev. X*, Vol. 2020.

# Supplementary files: Berreman Embedded Eigenstates for Narrowband Absorption and Thermal Emission

Zarko Sakotic[1*], Alex Krasnok[2], Norbert Cselyuszka[1,3], Nikolina Jankovic[1], and Andrea Alú[2*]

[1]BioSense Institute-Research Institute for Information Technologies in Biosystems, University of Novi Sad, Dr Zorana Djindjica 1a, 21101, Novi Sad, Serbia

[2]1Advanced Science Research Center, City University of New York, New York, NY 10031, USA

[3]Silicon Austria Labs, Sensor Systems, Microsystem Technologies, Villach, Austria

*To whom correspondence should be addressed: zarko.sakotic@biosense.rs, aalu@gc.cuny.edu

### 1. Mode dispersion and the reflection coefficient

As discussed in the main text, we use the standard transverse resonance (TR) technique to obtain the mode dispersion $Z_{\text{up}} + Z_{\text{down}} = 0$, [1,2]. For the single slab of ENZ (figure S1), the TR condition at the center of the structure yields:

$$Z_{\text{up}} + Z_{\text{down}} = \frac{2Z_1\left(Z_0 + jZ_1 \tan\left(\frac{k_{1z}t}{2}\right)\right)}{Z_1 + jZ_0 \tan\left(\frac{k_{1z}t}{2}\right)} = 0.$$

Setting the numerator to zero leads to:

$$\tan\left(\frac{k_{1z}t}{2}\right) = j\frac{Z_0}{Z_1} = j\frac{k_{0z}\varepsilon_{enz}}{k_{1z}},$$

where $k_{0z} = \sqrt{k_0^2 - k_x^2}$, $k_{1z} = \sqrt{k_0^2 \varepsilon_{enz} - k_x^2}$ are propagation constants along z-direction in air and ENZ respectively, $Z_0 = k_{0z}/\omega\varepsilon_0$ and $Z_1 = k_{0z}/\omega\varepsilon_0\varepsilon_{enz}$ are transverse magnetic (TM) wave impedances of air and ENZ, and *t* is the slab thickness. This dispersion relation corresponds to the even mode dispersion. The odd mode TM dispersion is obtained by placing an open circuit in the center of the transverse network, which acts as a magnetic wall. This gives the dispersion relation

$$\cot\left(\frac{k_{1z}t}{2}\right) = j\frac{Z_0}{Z_1} = j\frac{k_{0z}\varepsilon_{enz}}{k_{1z}}.$$

The same condition for the 3 layer structure yields

$$Z_{\text{up}} + Z_{\text{down}} = 2Z_2 \frac{Z_1(Z_0 + jZ_1\tan(k_{1z}t)) + jZ_2\tan\left(k_{2z}\frac{d}{2}\right)(Z_1 + jZ_0\tan(k_{1z}t))}{Z_2(Z_1 + jZ_0\tan(k_{1z}t)) + jZ_1\tan\left(k_{2z}\frac{d}{2}\right)(Z_0 + jZ_1\tan(k_{1z}t))} = 0.$$

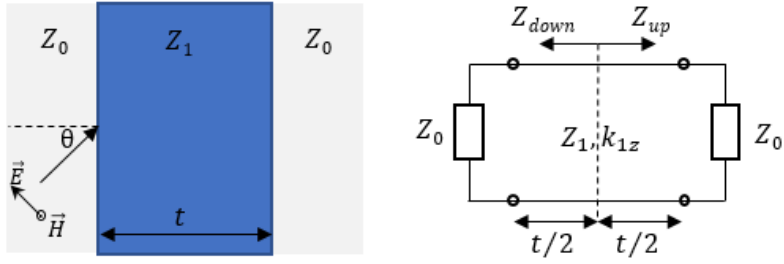

Figure S1. Transverse resonance network model for the single slab of ENZ.

Setting the numerator to zero leads to the even mode dispersion:

$$\tan\left(\frac{k_{2z}d}{2}\right) = j\frac{Z_1}{Z_2}\frac{(Z_0 + jZ_1\tan(k_{1z}t))}{(Z_1 + jZ_0\tan(k_{1z}t))} = j\frac{k_{1z}\varepsilon_2}{k_{2z}\varepsilon_1}\frac{(k_{0z} + j\frac{k_{1z}}{\varepsilon_{enz}}\tan(k_{1z}t))}{(\frac{k_{1z}}{\varepsilon_{enz}} + jk_{0z}\tan(k_{1z}t))}.$$

By putting an open circuit in the center of the transverse network, we obtain the odd mode dispersion:

$$\cot\left(\frac{k_{2z}d}{2}\right) = j\frac{Z_1}{Z_2}\frac{(Z_0 + jZ_1\tan(k_{1z}t))}{(Z_1 + jZ_0\tan(k_{1z}t))} = j\frac{k_{1z}\varepsilon_2}{k_{2z}\varepsilon_1}\frac{(k_{0z} + j\frac{k_{1z}}{\varepsilon_{enz}}\tan(k_{1z}t))}{(\frac{k_{1z}}{\varepsilon_{enz}} + jk_{0z}\tan(k_{1z}t))}.$$

Solutions for these relations are found within the parameter space consisting of complex $\omega$ and real $k_x$.

Throughout the main text, we use the reflection coefficient for single-layer and multilayer structures. This is obtained using the standard ABCD matrix formulation, where each layer can be represented by an equivalent ABCD matrix [3,4]:

$$\begin{bmatrix} A_n & B_n \\ C_n & D_n \end{bmatrix} = \begin{bmatrix} \cos(k_{nz}t_n) & jZ_n\sin(k_{nz}t_n) \\ \frac{j}{Z_n}\sin(k_{nz}t_n) & \cos(k_{nz}t_n) \end{bmatrix}.$$

where $n$ denotes the layer in question. For a single slab this yields the reflection coefficient:

$$r_1 = \frac{A + \frac{B}{Z_0} - CZ_0 - D}{A + \frac{B}{Z_0} + CZ_0 + D} = \frac{j\frac{Z_1^2 - Z_0^2}{2Z_1Z_0}\sin(k_{1z}t)}{\cos(k_{1z}t) + j\frac{Z_1^2 + Z_0^2}{2Z_1Z_0}\sin(k_{1z}t)}$$

When it comes to TM wave impedances, they can be expressed in terms of angles as:

$$Z_0 = \eta_0\cos\theta_i, \qquad Z_1 = \frac{\eta_0\cos\theta_1}{\sqrt{\varepsilon_{enz}}},$$

where $\theta_i$ is the incident angle, $\theta_1$ is the angle of refraction in the ENZ layer which is obtained using Snell's law. Then we can express the reflection coefficient in terms of angles:

$$r_1 = \frac{j\sqrt{\varepsilon_{enz}}\dfrac{\dfrac{\cos^2\theta_1}{\varepsilon_{enz}} - \cos^2\theta_i}{2\cos\theta_1\cos\theta_i}\sin(k_{1z}t)}{\cos(k_{1z}t) + j\sqrt{\varepsilon_{enz}}\dfrac{\dfrac{\cos^2\theta_1}{\varepsilon_{enz}} + \cos^2\theta_i}{2\cos\theta_1\cos\theta_i}\sin(k_{1z}t)}.$$

For the three-layer structure the total ABCD matrix is:

$$\begin{bmatrix} A_t & B_t \\ C_t & D_t \end{bmatrix} = \begin{bmatrix} A_1 & B_1 \\ C_1 & D_1 \end{bmatrix}\begin{bmatrix} A_2 & B_2 \\ C_2 & D_2 \end{bmatrix}\begin{bmatrix} A_3 & B_3 \\ C_3 & D_3 \end{bmatrix}$$

and the reflection coefficient for the three-layer structure is given by:

$$r_3 = \frac{A_t + \dfrac{B_t}{Z_0} - C_t Z_0 - D_t}{A_t + \dfrac{B_t}{Z_0} + C_t Z_0 + D_t}.$$

2. **Brewster's condition derivation**

As mentioned in the main text, the only possible zero of reflection for the single slab of ENZ material occurs due to non-resonant impedance matching (TM waves). Here we derive the underlying Brewster's condition. Namely, due to matched TM wave impedances, we can write:

$$Z_0 = \eta_0 \cos\theta_i = Z_1 = \eta_0 \cos\theta_1 \frac{1}{\sqrt{\varepsilon_{enz}}}, \quad \eta_0 = \sqrt{\frac{\mu_0}{\varepsilon_0}},$$

$$\cos\theta_1 = \sqrt{\varepsilon_{enz}}\cos\theta_i,$$

where $\theta_i$ and $\theta_1$ correspond to wave angles in air and ENZ, respectively. Snell's law for the air-ENZ boundary states:

$$\sin\theta_1 = \frac{1}{\sqrt{\varepsilon_{enz}}}\sin\theta_i.$$

By squaring the previous two equations and adding them together, the system of equations leads to the well known Brewster's formula:

$$\sin^2\theta_1 + \cos^2\theta_1 = 1 = \sin^2\theta_i + \cos^2\theta_i = \frac{1}{\varepsilon_{enz}}\sin^2\theta_i + \varepsilon_{enz}\cos^2\theta_i$$

$$\frac{(\varepsilon_{enz} - 1)}{\varepsilon_{enz}}\sin^2\theta_i = (\varepsilon_{enz} - 1)\cos^2\theta_i$$

$$\tan\theta_i = \tan\theta_B = \sqrt{\varepsilon_{enz}}$$

This leads to the equation (4) from the main text:

$$\theta_B = \text{atan}(\sqrt{\varepsilon_{enz}}) = \text{atan}\left(\sqrt{1-\left(\frac{\omega_p}{\omega}\right)^2}\right), \omega \geq \omega_p.$$

### 3. Berreman mode: analogy to a lumped circuit element

Here we present the analogy between the three-layer structure from the main text and a lumped element circuit, which gives the Berreman mode a simple and intuitive explanation. First, we start by replacing the ENZ layers by a series inductor, as in Figure S2 (b). Namely, just below plasma frequency the permittivity of the ENZ slab is negative and there is no phase advance through the slab since the wave in the slab is purely evanescent, thus allowing us to replace it by an inductor.

We choose a fixed frequency value below plasma frequency $\omega = 0.98\omega_p$ where this analogy will be valid, which gives $\varepsilon_{enz} \approx -0.04$. Next, we choose an arbitrary input angle, e.g. $\theta = \sin^{-1}(k_x/k_0) = 50°$. For the inductor, we choose an appropriate reactance value $X_L = \omega L$ to accurately portray the ENZ slab in these conditions. For these parameters we found that value to be $X_L = 2270\,\Omega$. To validate this analogy, we plot the reflection coefficient and input impedance for the original structure [Figure S2 (a), same as Figure 2 (b) in the main text], and the transmission line model with inductors [Figure S2 (b)], all as a function of the dielectric thickness $d$.

We now vary the dielectric gap thickness and compare the results. Figure S2 (c) shows an excellent agreement between the two reflection coefficients, where $0^{th}$ and $1^{st}$ order Berreman modes (reflection zeros) are visible. Figure S2 (d) and S2 (e) show excellent agreement in trend and value between the real and imaginary parts of the input impedance, making this analogy appropriate.

Furthermore, if we only look at the $0^{th}$ order Berreman mode which exists for a very small $d$, we can assume $k_{2z}d \ll 1$ (small phase advance) and replace the dielectric gap by a shunt capacitor, Figure S2 (f). In this case, we can mimic the change of dielectric thickness to some extent by changing the capacitance $C$, effectively changing the reactance $X_C$.

We now plot the reflection coefficient and input impedance of *L-C-L* circuit (Figure S2 (g), (h) and (i)) as a function of capacitance $C$. One can notice there are now two *x*-axis, with each curve corresponding to its own *x*-axis. We compare the mentioned parameters of *L-C-L* circuit to that of the original structure, which are a function of thickness $d$. These parameters show the same trends, as the dielectric thickness $d$ (bottom, blue *x*-axis) and capacitance $C$ (top, red *x*-axis) linearly change. However, as the thickness $d$ increases, the propagation effects in the dielectric become noticeable, and there is a discrepancy in terms of the position of the reflection dips. Nevertheless, this model successfully provides an intuitive picture of the underlying physics, as the analogous circuit accurately mimics the resonant behavior of the multilayer structure.

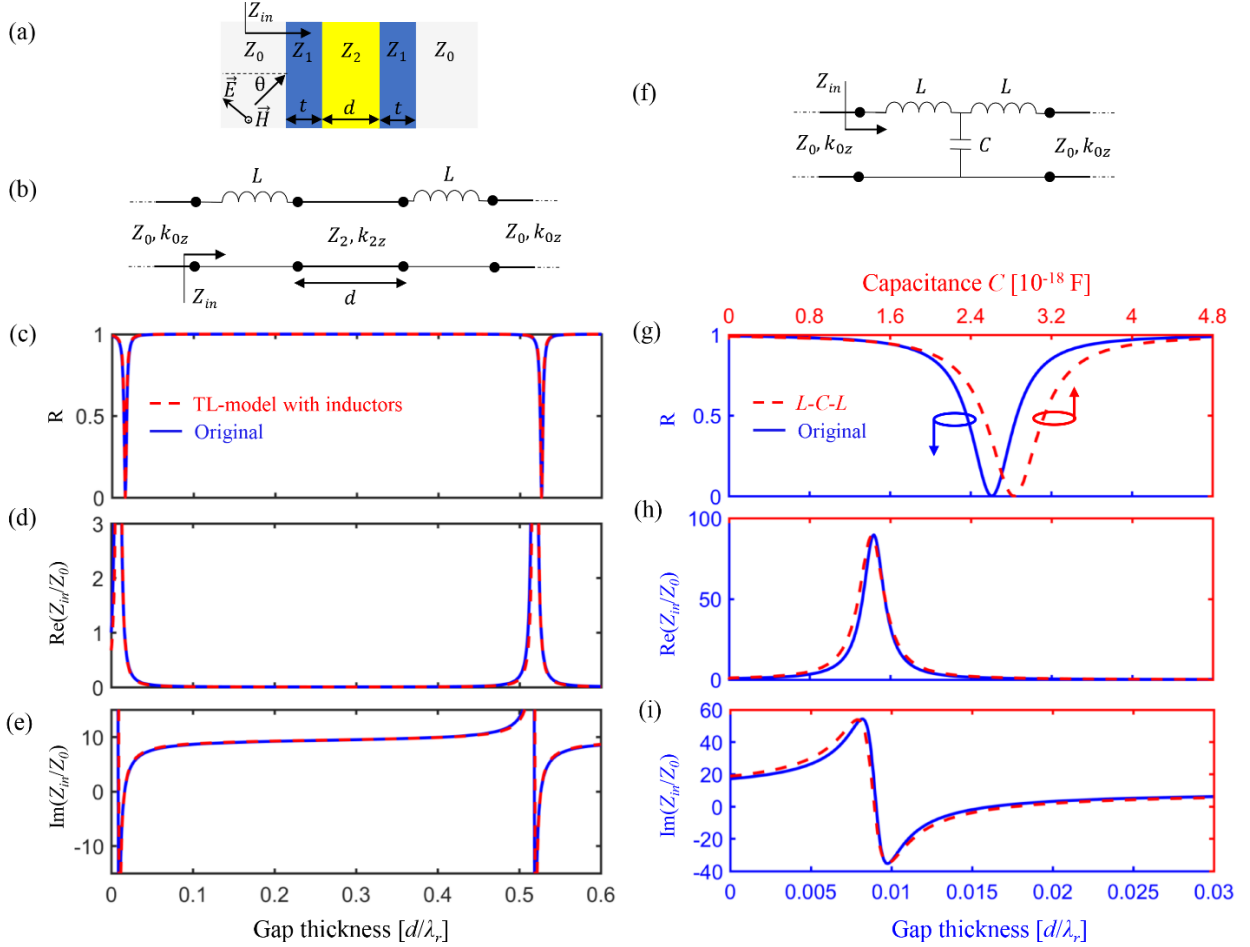

Figure S2. (a) Three-layer structure from the main text. (b) Transmission line model of the structure, ENZ replaced by an inductor (c) Reflection coefficient of (a) and (b). (d) Real part of the input impedance of (a) and (b). (e) Imaginary part of the input impedance of (a) and (b), (f) Transmission line model of the *L-C-L* circuit, (g) Reflection coefficient of (a) and (f), as a function of $d$ and $C$, respectively. (h) Real part of the input impedance of (a) and (f), (i) Imaginary part of the input impedance of (a) and (f).

Thus, it is clear that the tunneling at the $0^{th}$ Berreman mode can be attributed to classical *R-L-C* resonance, where reactance contributions of ENZ and dielectric layers balance out, allowing full energy transmission with negligible phase advance.

## 4. Symmetry protected embedded eigenstates

In terms of S-matrix eigenvalues, a single slab of ENZ material has a single complex pole and a single complex zero, as well as a reflection zero on the real frequency axis. Figure S3 (a) shows the trajectories of the singularities as $k_x$ reduces to 0, where all these singularities merge and form an embedded eigenstate.

In the case of ENZ-dielectric-ENZ for the same incidence angle, two poles and two zeros are clearly present, Figure S3 (b) and (c). In both cases, the poles and zeros collapse to the real frequency $\omega_p$ for $k_x=0$, creating a doubly degenerate embedded eigenstate.

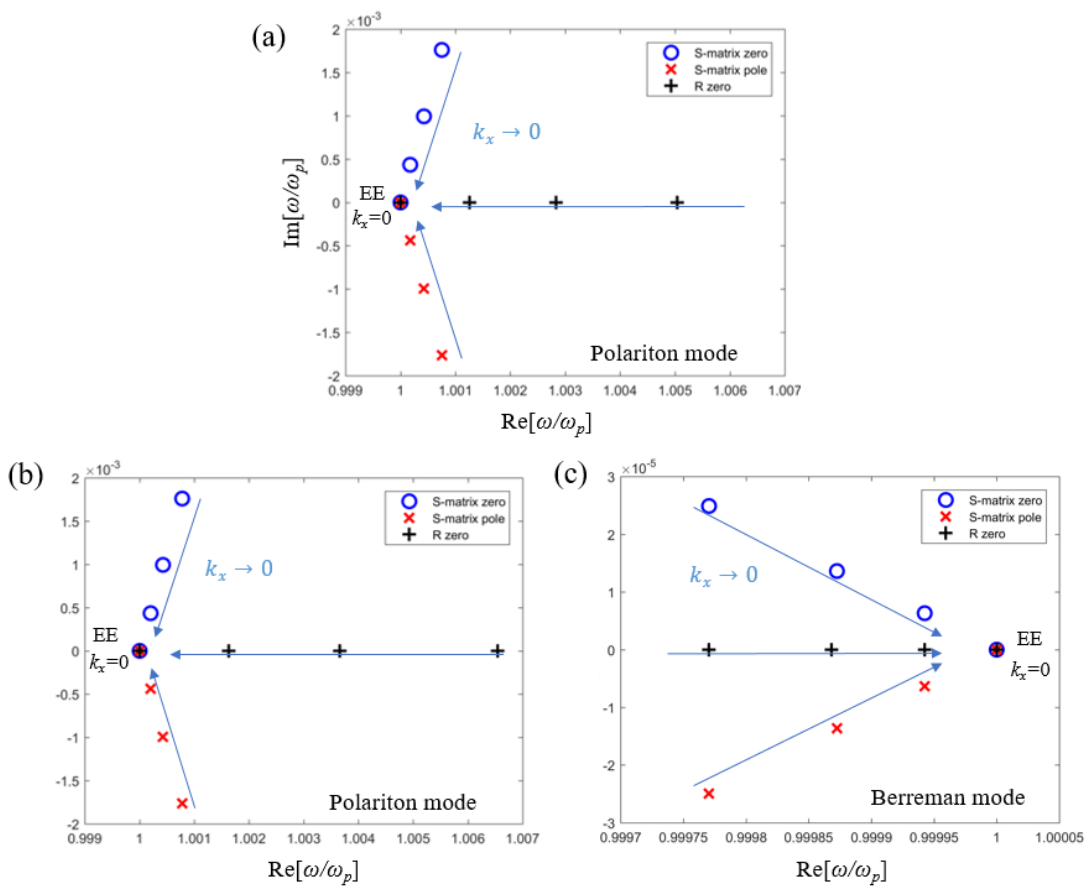

Figure S3. Singularities of the S-matrix eigenvalues/reflection spectrum in the complex frequency plane near normal incidence for 4 different values of $k_x/k_p$ = [0.1, 0.075, 0.05, 0]: (a) Single slab of ENZ of thickness $t$=800 nm, (b) Polariton mode singularities for ENZ-dielectric-ENZ, $t$=400 nm, and the dielectric thickness $d$=20 nm. (c) Berreman mode for (b).

### 5. Polariton mode in a free-standing ENZ slab

The use of a SiC-resonator-SiC configuration was demonstrated in the main text. Here we show that the absorption properties of a single SiC slab that supports a polariton mode are inherently inferior to that of the proposed configuration. Namely, by tuning the thickness of a single slab of SiC, Figure S5 shows that the Berreman mode is inherently flat (no dispersion control), has poor angular selectivity, and does not admit perfect absorption.

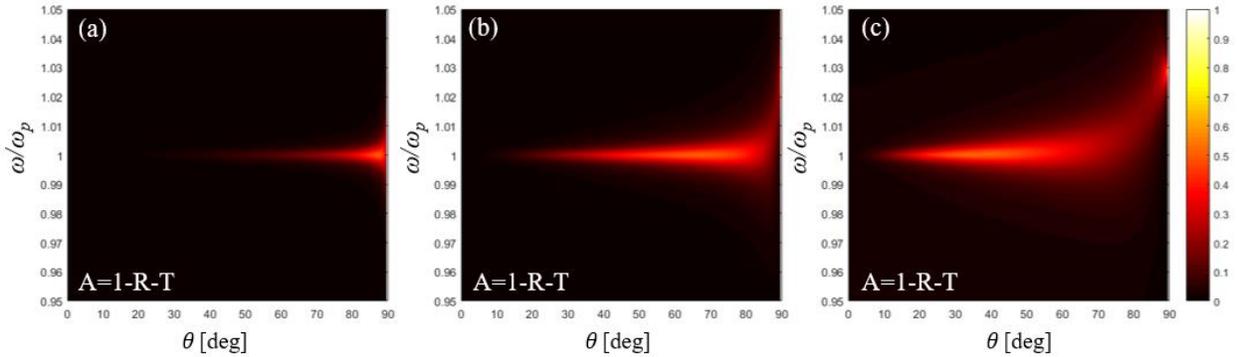

Figure S4. TM absorption coefficient for a single slab of SiC with thickness t (a) *t*=10 nm. (b) *t*=100 nm. (c) *t*=500 nm.

**References**

[1]  F. Monticone and A. Alù, "Leaky-Wave Theory, Techniques, and Applications: From Microwaves to Visible Frequencies," Proc. IEEE, vol. 103, no. 5, pp. 793–821, May 2015.

[2]  N. Marcuvitz, "On field representations in terms of leaky modes or Eigenmodes," IRE Trans. Antennas Propag., vol. 4, no. 3, pp. 192–194, Jul. 1956.

[3]  D. M. Pozar, Microwave engineering, 4th Edition. John Wiley & Sons, Inc., 2011.

[4]  Y. Li and C. Argyropoulos, "Exceptional points and spectral singularities in active epsilon-near-zero plasmonic waveguides," Phys. Rev. B, vol. 99, no. 7, pp. 1–43, 2019.